# Netizens, Academicians, and Information Professionals' Opinions About AI With Special Reference To ChatGPT

Subaveerapandiyan A

Vinoth A

Neelam Tiwary

# Netizens, Academicians, and Information Professionals' Opinions About AI With Special Reference To ChatGPT

**Subaveerapandiyan A**

Ph.D. Research Scholar

Department of Library and Information Science

Yenepoya (Deemed to be University), Mangalore, Karnataka, India

Email: subaveerapandiyan@gmail.com

ORCiD: https://orcid.org/0000-0002-2149-9897

**Vinoth A**

Ph.D. Research Scholar

School of Computer Science and Information Technology

DMI-St. Eugene University, Lusaka, Zambia

Email: jeba.vinot@gmail.com

**Neelam Tiwary**

Ph.D. Research Scholar

School of Computer Science and Information Technology

DMI-St. Eugene University, Lusaka, Zambia

E-mail: neelamtiwary06@gmail.com

**Abstract**

This study aims to understand the perceptions and opinions of academicians towards ChatGPT-3 by collecting and analyzing social media comments, and a survey was conducted with library and information science professionals. The research uses a content analysis method and finds that while ChatGPT-3 can be a valuable tool for research and writing, it is not 100% accurate and should be cross-checked. The study also finds that while some academicians may not accept ChatGPT-3, most are starting to accept it. The study is beneficial for academicians, content developers, and librarians.

**Keywords:** Conversational Generative Pre-training Transformer (ChatGPT), Artificial Intelligence in Academia, Academic Writing with ChatGPT, Library Services

**Introduction**

The OpenAI-developed GPT (Generative Pre-trained Transformer) model has a variation called ChatGPT. The GPT model was initially released in 2018 and trained using the Common Crawl, a sizable dataset of text from the internet. The Transformer design, revealed in a 2017 study by Google researchers, served as the model's foundation. Unsupervised learning was used to train the initial GPT model, which meant that it was trained on a sizable text dataset without any explicit labels or annotations. As a result, the model could pick up on various textual patterns and structures and produce new text with a similar tone and structure (Wikipedia 2023).

A private artificial intelligence research facility made up of the for-profit OpenAI LP, and its parent organization, the nonprofit OpenAI Inc., ChatGPT, is owned and developed by OpenAI. It is also known as GPT-3 (Generative Pre-trained Transformer 3). Elon Musk, Sam Altman, Greg Brockman, Ilya Sutskever, Wojciech Zaremba, and several others created OpenAI in December 2015 to advance and advance benign AI in a way that benefits humankind as a whole. San Francisco, California, USA, is home to the business (Wikipedia 2023; "About OpenAI" 2015).

ChatGPT, also known as GPT-3 (Generative Pre-trained Transformer 3), is the latest version of a series of language models developed by OpenAI.

***The previous versions of GPT include:***

- ➢ GPT (Generative Pre-trained Transformer): The first version of GPT was introduced in 2018; it was trained on a massive dataset of text from the internet and could generate human-like text. GPT-2 (Generative Pre-trained Transformer
- ➢ The second version of GPT, GPT-2, was released in 2019; it was an even larger model with over 1.5 billion parameters and could generate text that was often indistinguishable from text written by humans.
- ➢ GPT-3 (Generative Pre-trained Transformer 3), also known as ChatGPT: The latest and the most advanced version of the GPT series, GPT-3 was introduced in June 2020, has 175 billion parameters, making it one of the largest and most powerful language models to date. It is trained on a diverse internet text and can perform a wide range of natural language processing tasks.

Since its introduction, GPT-3 has gained significant traction and is now utilized in various fields, including language-based games, chatbots, virtual assistants, language production, and translation. It has won recognition for its capacity to produce writing that resembles that of a human being and to carry out various natural language processing tasks. ChatGPT can be a helpful tool for text generation and comprehending natural language, but it is essential to utilize it carefully and double-check the data it generates.

**Features and Future of ChatGPT**

Usage The API enables programmers, academics, and data scientists to incorporate the capabilities of GPT-3 into their software and systems, including

- *Chatbots and virtual assistants:* ChatGPT can build chatbots and virtual assistants that comprehend user input and respond conversationally and naturally. Chatbots can be a cost-effective approach to responding to most common reference inquiries and guiding customers to the appropriate services. Still, they cannot match the complexity of human interaction (both intellectually and emotionally). Chatbots can forge a closer connection with the young generation. Chatbots can instantly and consistently respond to numerous users' questions that are being asked at the same time. (McNeal and Newyear 2013; Vincze 2017).

- *Language-based games:* ChatGPT can create language-based games and interactive fiction, where the model generates responses to user input in real time. Learning environments, knowledge-sharing platforms, and instruments for knowledge transmission all exist within video games. Researchers contend that libraries can close the digital divide by fostering a space for digital video games (Gee 2012).

- *Language translation:* Before the invention of language translation tools, libraries gave language translation services to the users, but now technologies are replacing translation jobs. The provision of translation services by libraries is crucial to improving the utilization of documents. When using the translation service, a library user typically asks the library to translate a needed article written in a language they do not understand. ChatGPT can be fine-tuned to translate text from one language to another, making it an effective tool for multilingual communication.

- *Content creation:* ChatGPT can produce a wide range of content, including text summaries, complete articles, and natural language answers to questions.

- ❖ *Libraries:* A few libraries have utilized ChatGPT to provide reports, automate library user service, and create content.
- ❖ *Content generation:* ChatGPT may create fresh, original content quickly and easily for websites, social media, and marketing materials.
- ❖ *Library Developers:* ChatGPT enables developers to create natural language processing applications like chatbots, virtual assistants, and question-and-answer systems.
- ❖ *Researchers:* Researchers can use ChatGPT to perform natural language understanding and generation tasks, such as text summarization and text completion. It can also help researchers by creating hypotheses, data summaries, and other things.
- ❖ *Educators and Students:* ChatGPT can be used by teachers and students to create summaries and essays, complete language translation and comprehension tasks, provide instructional content, tests, and questionnaires, and assist students with their assignments.
- ❖ *Writers and content creators:* Writers and content producers can utilize ChatGPT to generate text, summaries, and ideas.
- ❖ *Government and non-profit organizations:* Government agencies and nonprofits can use ChatGPT to complete activities, including document summarising, language translation, and automated content creation.
- ❖ *Data Scientists:* In order to generate and process a massive volume of text data for usage in various data science projects and analyses, ChatGPT can be employed.
- ❖ *Media companies and publishers:* ChatGPT can be used to create text summaries and content, help with editing, and more.
- ❖ *Individual users:* utilize ChatGPT to help them write emails, documents, and other text-based content. Text classification and sentiment analysis: The text classification and sentiment analysis capabilities of ChatGPT can be honed for usage in applications like social media monitoring and customer feedback analysis.
- ❖ *Text-to-speech:* To provide verbal responses, ChatGPT can be connected with text-to-speech technology, making it useful for chatbots and interactive voice assistants.

Overall, ChatGPT is a flexible tool that may be used by various users and in various applications. However, as was previously stated, it is crucial to use it carefully and double-check the information it offers.

***In academic writing, there are several potential issues associated with using ChatGPT:***

- ***Reduced critical thinking:*** Adopting language models like ChatGPT can lead to a decrease in critical thinking abilities. This is because the model can rapidly and simply deliver information and respond to inquiries, potentially eliminating the need for people to engage in independent critical thought and problem-solving. Furthermore, if the model's information is reliable and accurate, it can cause people to believe erroneous information without checking it.
- ***Ethical concerns:*** AI in academic writing raises ethical concerns about using automated tools to produce work that is then attributed to human authors. There are several ethical concerns related to using AI in academic writing, specifically language models such as ChatGPT. Some of these concerns include the following:

    *i) Transparency:* As mentioned earlier, using language models in academic writing can raise concerns about transparency and accountability. It needs to be clarified to readers how much of the content was generated by the model and how much was written by the researcher.

    *ii) Plagiarism:* Using language models to generate text can make it easier for researchers to engage in plagiarism by presenting text generated by the model as their work. AI-powered writing tools can make plagiarism-free text or less plagiarism by paraphrasing.

    *iii) Bias:* Language models can be trained on partial data, which can perpetuate bias in the generated text. AI-powered tools used in academic research may perpetuate biases in the data they are trained on, leading to unsatisfactory or discriminatory results. Bias in research refers to the systematic error in a study's design, conduct, or analysis, which can lead to incorrect conclusions. Bias can occur in various forms, including selection bias, measurement bias, and publication bias.

    *iv) Loss of creativity:* Using language models to generate text can lead to decreased creativity and critical thinking among researchers, who may rely on the model to generate ideas and text. AI-generated text may be complex for humans to understand, making it difficult to evaluate the quality and accuracy of the work produced.

*v) Misuse:* Language models can fabricate data or results, which could lead to fraudulent research.

*vi) Misuse of the technology:* AI-generated text can be used to create fake news and misleading information, which can be spread through academic circles and cause harm to the reputation of the researcher or the institution. Additionally, using AI in academic writing raises concerns about the impact on jobs and the future of work. As AI-generated content becomes more sophisticated, it may replace the need for human writers in specific fields, leading to job losses and economic disruption. It is vital for researchers, academic institutions, and publishers to be aware of these ethical concerns and to establish guidelines and policies for using AI in academic writing. It includes ensuring transparency and accountability, evaluating the potential for bias, and considering the impact on jobs and the economy. AI has the potential to revolutionize academic writing. Still, it also poses new challenges that must be addressed to ensure the technology is used responsibly and ethically.

AI cannot think systematically like humans, which is necessary to create a multi-paragraph essay that calls for exact and thorough data (*The Guardian* 2020). Research support is one of ChatGPT-3's critical benefits in academic writing, as it can assist in tasks such as summarizing documents, highlighting important ideas, and providing citations. It can save time and effort for researchers, allowing them to focus on more important activities such as analysis and interpretation. ChatGPT-3 can also aid the writing process by generating text for academic papers such as research papers, essays, and dissertations. Furthermore, it can offer advice on grammar, style, and coherence to help writers improve their writing (Aljanabi et al. 2023).

*Can AI replace the librarian?*

AI technology could automate some of the traditionally performed by librarians, such as cataloging and information retrieval. However, it is essential to note that librarians play a critical role in many aspects of library operations beyond these tasks. One of the critical roles of librarians is to help users find and use information effectively. They provide expert guidance and instruction and curate and organize information resources. These tasks are often complex and require a high level of knowledge and expertise, which may be difficult for AI systems to replicate.

Additionally, Librarians are responsible for preserving the library collections and maintaining the integrity of the information resources. This is a task that AI cannot replace. Librarians also play an essential role in the community, engaging patrons and providing them with access to information and resources that may not be available online. AI technology can automate some tasks traditionally performed by librarians. However, it is unlikely to fully replace the role of librarians in providing expert guidance, literature searches, curating information resources, cataloging and indexing books, virtual tours, managing library collections, and assisting patrons with research and engaging with patrons. Librarians have unique skills and knowledge, including expertise in information organization, research methods, and copyright laws. Librarians often have specialized knowledge and training in specific areas, such as archival studies or rare books, which would be difficult for an AI system to replicate. They also provide valuable services such as reference assistance, reader advisory, and programming.

**Research Method**

The research methodology employed in the study aimed to gather both quantitative and qualitative data to analyze social media comments and content related to ChatGPT. The qualitative data was collected from various social media platforms, including LinkedIn, Facebook, and Twitter. The researcher selected only the relevant comments that were informative and worth analyzing and discussing, including both positive and negative views on the use of ChatGPT in research and writing.

To gather quantitative data, the researcher utilized six questionnaires and sent them to library professionals via WhatsApp groups. The data collection took place between January 15th to 31st, 2023, with 71 respondents. To analyse qualitative data, five questions are used.

In conclusion, the research methodology employed in this study was designed to provide a comprehensive understanding of the views and opinions on ChatGPT by examining social media comments and content. The combination of qualitative and quantitative data allowed for a more thorough examination of the topic and a more in-depth analysis of the results.

**Data Analysis**

## Part - I Quantitative Analysis

Table 1. Do you know What ChatGPT is?

| Awareness of ChatGPT | Respondent | Percentage |
|---|---|---|
| Yes | 68 | 95.8 |
| No | 3 | 4.2 |

This table 1 represents the results of a survey asking respondents about their awareness of ChatGPT. 68 respondents (95.8%) answered "Yes," indicating that they know what ChatGPT is, while 3 respondents (4.2%) answered "No," indicating that they do not know what it is.

Table 2. Have you used ChatGPT?

| Experience usage of ChatGPT | Respondent | Percentage |
|---|---|---|
| Yes | 65 | 91.5 |
| No | 6 | 8.5 |

Table 2 represents the results of a survey asking respondents about their experience using ChatGPT. 65 respondents (91.5%) answered "Yes," indicating that they have used ChatGPT, while 6 respondents (8.5%) answered "No," indicating that they have not used it.

Table 3. If so, what purpose did you use ChatGPT for?

| Purpose of used | Respondent | Percentage (N=7) |
|---|---|---|
| Language correction | 44 | 62 |
| Sentence making | 34 | 47.9 |
| Paraphrasing | 18 | 25.4 |
| General knowledge | 45 | 63.4 |
| Searching, browsing and surfing | 58 | 81.7 |
| Grammar corrections | 48 | 67.6 |
| Summarization | 39 | 54.9 |
| Letter preparing | 37 | 52.1 |
| Quick answers finding | 59 | 83.1 |
| Research | 19 | 26.8 |
| Data Analysis | 14 | 19.7 |
| Conversation | 33 | 46.5 |

| Language Translation | 15 | 21.1 |
| None of the above | 14 | 19.7 |

Table 3 shows the results of a survey asking respondents about the purpose for which they used ChatGPT. The total number of respondents who used ChatGPT is 71. Some of the most common uses for ChatGPT include language correction (62%), sentence making (47.9%), searching, browsing, and surfing (81.7%), and quick answers finding (83.1%). Other uses listed in the table include paraphrasing, general knowledge, grammar corrections, summarization, letter preparation, research, data analysis, conversation, language translation, and none of the above.

Table 4. Do you think ChatGPT is against academic integrity?

| ChatGPT is against academic integrity | Respondent | Percentage |
| --- | --- | --- |
| Yes | 24 | 33.8 |
| No | 47 | 66.2 |

Table 4 reveals the survey results asking respondents about their opinions on whether they think ChatGPT is against academic integrity. 24 respondents (33.8%) answered "Yes," indicating that they believe ChatGPT is against academic integrity, while 47 respondents (66.2%) answered "No," indicating that they do not believe it is against academic integrity.

Table 5. Will you adopt ChatGPT for Library Services?

| Will you adopt ChatGPT for Library Services? | Respondent | Percentage |
| --- | --- | --- |
| Yes | 45 | 63.4 |
| No | 16 | 22.5 |
| Not sure | 10 | 14.1 |

Table 5 shows the results of a survey asking respondents about their willingness to adopt ChatGPT for library services. 45 respondents (63.4%) answered "Yes," indicating that they would adopt ChatGPT for library services, while 16 respondents (22.5%) answered "No," indicating that they would not adopt it. 10 respondents (14.1%) answered "Not sure," indicating they are uncertain about adopting ChatGPT for library services.

Table 6. Will you do the subscription ChatGPT for Library Services?

| Will you subscribe to ChatGPT | Respondent | Percentage |
| --- | --- | --- |
| Yes | 31 | 43.7 |
| No | 18 | 25.4 |
| Not sure | 22 | 30.9 |

Table 6 resembles the results of a survey asking respondents about their willingness to subscribe to ChatGPT for library services. 31 respondents (43.7%) answered "Yes," indicating that they would subscribe to ChatGPT for library services, while 18 respondents (25.4%) answered "No," indicating that they would not subscribe. 22 respondents (30.9%) answered "Not sure," indicating that they are uncertain about subscribing to ChatGPT for library services.

## Part - II Content Analysis

*What exactly is ChatGPT?*

*Comments*

"My new assistant's name"

"Chatting platform"

A well-developed oligopoly manages newspeak with a deceptive term that appears transparent on the surface. Limiting narratives, freethought, and analytical thinking to amass an infinite amount of data creates an intellectual monopoly. It seeks to replace the working class with a less demanding and more productive substitute. It could be the solution to and the root of all of humanity's problems in the future. Whether we like it or not, the future has arrived.

*Interpretation*

ChatGPT is an excellent tool for generating ideas and providing suggestions. However, it is not a replacement for the creativity, critical thinking, and audience understanding essential for producing high-quality written content. ChatGPT uses an unsupervised learning technique, which does not require explicit instructions or labeled data for training. Instead, it learns by analyzing patterns and relationships in the data it has been trained on.

*Do we need to prohibit ChatGPT?*

*Comments*

New tools and technologies can be beneficial as long as they are used to augment and not replace our ability to think and create. It can become a concern when we give up our ability and become dependent on technology. Therefore, it is vital to use tools such as ChatGPT-3 to aid research and writing rather than relying solely on them to generate ideas and text. Using it this way, we can take advantage of its capabilities while still maintaining our creativity and independence.

It is vital to embrace new tools and technologies, such as ChatGPT-3, as long as they are used to augment and not replace our cognitive abilities. Resisting new technologies out of fear or Luddism will not help us to develop and improve our skills. Instead, educators should focus on developing new learning environments that incorporate these technologies and challenge us to use them in a meaningful and productive way.

Banning ChatGPT, or any other technology for that matter is not a productive solution as it is just another step in the evolution of technology. Just like the invention of Google or calculators, ChatGPT is a tool that can enhance our abilities and make tasks more efficient. The key is to use these technologies to augment our cognitive abilities and skills rather than replace them. Instead of banning ChatGPT, educating people on how to use it responsibly and effectively would be more beneficial.

Banning ChatGPT or any other technology is not the solution, as it is a new and constantly evolving tool. ChatGPT-4 will likely be even more advanced, and understandably, it may take some time to understand and use it fully. Instead of banning this tool or any other technology, it would be more beneficial to take the time to learn and understand its capabilities and limitations and use it responsibly and effectively. It is important to note that many tools, such as Grammarly and Google predictive text, are AI-based technologies and may pose similar questions. It is vital to approach all AI-based tools with a critical and informed mindset.

It is possible that an over-reliance on technology could lead to a decline in specific skills, such as mental math. However, technology can also serve as a tool to enhance and augment human abilities rather than replace them entirely. Additionally, as technology advances, new skills and abilities may become necessary to use and navigate it effectively.

*Interpretation*

It is clear from the statements that banning new technology such as ChatGPT is not the right solution, as it is a tool that can be used to enhance our abilities and make tasks more efficient. Rather than banning it, it would be more beneficial to educate people on how to use it responsibly and effectively. It is important to remember that technology such as ChatGPT is only a tool and cannot replace human thinking and creativity. We should use it as an aid, not a replacement for our cognitive abilities and skills. Embracing new technologies with a critical and informed mindset is vital to using their benefits while avoiding potential drawbacks.

*Numerous scientists disagree with ChatGPT's listing as an author in research articles (Stokel-Walker 2023).*

*Comments:*

It is not uncommon for scientists to disapprove of using AI-based tools such as ChatGPT being listed as an author on research papers. These tools cannot conduct original research or contribute to the scientific process as human researchers do. Additionally, some scientists may view the use of ChatGPT in this way as an attempt to circumvent traditional methods of authorship and give undue credit to the technology rather than the human researchers who conducted the work. It is essential to consider the ethics and implications of using AI-based tools in scientific research and ensure that authorship is attributed appropriately.

It cannot be a co-author on a research paper. However, it can assist in generating text, highlighting important ideas, and providing citations. So, it can be used as a reference tool rather than as a co-author. It is crucial to ensure that authorship is attributed appropriately and that the contributions of human researchers are recognized.

Nice to see the most in-demand ghostwriter on the first page.

**Can AI write like humans?**

*Comments*

ChatGPT, like any other AI-based language model, may require significant human editing to produce a high-quality abstract. The generated text by ChatGPT is based on the input it receives, but it may only sometimes provide accurate or complete information in the desired format. While ChatGPT can assist in generating text, it is ultimately the researcher's responsibility to ensure accuracy, coherence, and relevance. It is essential to review, edit, and revise the generated text to ensure that it effectively communicates the key findings and contributions of the research.

*Interpretation*

Technology is constantly evolving, and new tools and technologies will continue to be developed. However, it is essential to remember that human thinking and creativity are unique and cannot be replaced by machines. Humans have emotions and feelings that allow us to express ourselves in ways that machines cannot replicate. Artificial Intelligence (AI) and other technological tools can assist and enhance human abilities, but they cannot replace the human experience and perspective. AI can generate text or even come up with ideas, but the final decision of the context, style, and tone of the writing is in the human's hand. In creative writing, the human experience and emotions play a crucial role in the process. The ability to convey emotions and feelings through writing is one of the things that sets human writing apart from the machine-generated text. Therefore, AI and other technological tools can be valuable resources. However, it is important to remember that they are just tools and should be used to supplement and enhance human abilities rather than replace them.

*ChatGPT requires further instruction before being deemed intelligent*
*Comments*
Even though I cannot generate pictures with ChatGPT, it has aided me in finding answers to various tasks. Yes, it needs upgrading, but it is my new friend.

The knowledge cutoff for ChatGPT is set at 2021, which is a drawback. It is also unable to use the internet to browse. The developers should thus work in this area to ensure a more extensive knowledge base. However, it is a fantastic program in and of itself.

ChatGPT is responding to my requests with the appropriate information but cannot provide references for crediting the original writers.

The only aspect of Ai that is currently flawless is its ability to write with proper grammar; everything else will require Ai to consult an outside source, and it still needs to be perfect at doing this.

*Interpretation*
Humans have a natural tendency to constantly seek improvement and innovation, which often leads to developing new inventions. However, it is also true that people will often find faults or imperfections in these new inventions and continue to strive for further advancements.

**Findings of the study**

**Part - I**

1. Awareness of ChatGPT: The majority of respondents (95.8%) are aware of what ChatGPT is, while a small percentage (4.2%) are not aware.
2. Experience with ChatGPT: The majority of respondents (91.5%) have used ChatGPT, while a smaller percentage (8.5%) have not used it.
3. Purpose of using ChatGPT: The most common uses for ChatGPT among the 71 respondents who used it include language correction (62%), sentence making (47.9%), searching, browsing and surfing (81.7%), and quick answers finding (83.1%).
4. ChatGPT and Academic Integrity: Opinions were divided on whether ChatGPT is against academic integrity, with 33.8% of respondents believing it is and 66.2% believing it is not.
5. Adoption of ChatGPT for Library Services: The majority of respondents (63.4%) would adopt ChatGPT for library services, while 22.5% would not and 14.1% are unsure.
6. Subscription to ChatGPT for Library Services: 43.7% of respondents would subscribe to ChatGPT for library services, 25.4% would not subscribe, and 30.9% are unsure about subscribing.

**Part - II**

ChatGPT is a powerful tool for generating text but is not a replacement for human creativity, critical thinking, and audience understanding. Scientists may disagree with listing ChatGPT as an author in research articles, as it cannot conduct original research or contribute to the scientific process. It can be used as a reference tool but must ensure that authorship is attributed appropriately. AI-based models, including ChatGPT, may require significant human editing to produce high-quality text, and it is ultimately the responsibility of the researcher to ensure accuracy, coherence, and relevance.

**Conclusion**

ChatGPT can be a valuable tool for academicians to improve their language and sentence structure in their writing. However, soon ChatGPT is being used in a business model. OpenAI is currently offering a free trial period for the API, but once the trial period is over, usage of the API will require a subscription. In the future, educational institutions may subscribe to tools like ChatGPT and other academic tools such as grammar correction, paraphrasing, plagiarism checking, and data analysis tools. This is standard practice in the technology industry, where software and tools are offered as a subscription service. It is important to remember that while these tools can be helpful, they should be used responsibly and ethically. This means using them to assist and enhance one's writing and research, but not to plagiarize or present someone else's work as one's own. Additionally, it is crucial to fact-check the model's information and verify the generated text's authenticity before using it for any purpose.